\documentclass[prd,twocolumn,amsmath,amssymb]{revtex4_my}
\usepackage{graphicx}

\begin{document}
\title{\bf \boldmath Measurements of the observed cross sections for
$e^+e^- \to$ light hadrons at $\sqrt{s}=$ 3.773 and 3.650 GeV}
\author{
M.~Ablikim$^{1}$,              J.~Z.~Bai$^{1}$,               Y.~Ban$^{12}$,
X.~Cai$^{1}$,                  H.~F.~Chen$^{16}$,
H.~S.~Chen$^{1}$,              H.~X.~Chen$^{1}$,              J.~C.~Chen$^{1}$,
Jin~Chen$^{1}$,                Y.~B.~Chen$^{1}$
% ...
\\
\vspace{0.5cm}
(BES Collaboration)\\
\vspace{0.5cm}
{\it
$^{1}$ Institute of High Energy Physics, Beijing 100049, People's Republic of China\\
% ...
$^{12}$ Peking University, Beijing 100871, People's Republic of China\\
% ...
$^{16}$ University of Science and Technology of China, Hefei 230026, People's Republic of China\\
% ...
}}

\begin{abstract}
Using the data sets of 17.3 pb$^{-1}$ collected at $\sqrt{s}=$ 3.773 GeV and
6.5 pb$^{-1}$ collected at $\sqrt{s}=$ 3.650 GeV with the BESII detector at
the BEPC collider, we have measured the observed cross sections for
18 exclusive light hadron final states produced in $e^+e^-$ annihilation at the two
energy points.
\end{abstract}

\maketitle

\oddsidemargin  -0.2cm
\evensidemargin -0.2cm

\section{Introduction}
The $\psi(3770)$ resonance is the lowest mass charmonium resonance
above the open charm pair $D\bar D$ production threshold.
It is expected to decay almost entirely into pure $D\bar D$ \cite{delco}.
However, it is difficult to understand for a long time that the observed
cross section $\sigma_{D\bar D}^{\rm obs}$ for $D\bar D$ production at the
peak of $\psi(3770)$ resonance is less than the observed cross section
$\sigma_{\psi(3770)}^{\rm obs}$ for $\psi(3770)$ production \cite{rosner,rzhc}.
Recently, BES Collaboration directly measured the branching fraction
for $\psi(3770) \to {\rm non-}D\bar D$ to be $(16.1\pm1.6\pm5.7)\%$ \cite{brdd1}
and $(16.4\pm7.3\pm4.7)\%$ \cite{brdd2} with two different data samples and
different methods. These imply that the $\psi(3770)$
might substantially decay into ${\rm non-}D \bar D$ final states. However, the
existing results about exclusive ${\rm non-}D\bar D$ decays of $\psi(3770)$
from BES \cite{bai,ablikim,kskl,rhopi} and CLEO \cite{adam,adams,coans,huang,cronin,briere}
Collaborations can not explain the discrepancy between the
$\sigma_{D\bar D}^{\rm obs}$ and $\sigma_{\psi(3770)}^{\rm obs}$.
To understand the discrepancy, at the first step, one can directly compare the observed cross
sections for $e^+e^-\to$ light hadrons at the center-of-mass energies
below 3.660 GeV and at 3.773 GeV, removing the contributions from $J/\psi$
and $\psi(3686)$ decays due to ISR returns as well as the contributions from
$D \bar D$ decays. These cross sections provide useful information to
understand the mechanisms of the continuum light hadron production
and helpful information to understand the discrepancy between the
$\sigma_{D\bar D}^{\rm obs}$ and $\sigma_{\psi(3770)}^{\rm obs}$.

In this Letter, we report measurements of the observed cross sections for
the exclusive light hadron final states of $\phi\pi^0$, $\phi\eta$,
$2(\pi^+\pi^-)$, $K^+K^-\pi^+\pi^-$, $2(K^+K^-)$, $p\bar p\pi^+\pi^-$,
$p\bar p K^+K^-$, $3(\pi^+\pi^-)$, $\phi\pi^+\pi^-$, $\phi K^+K^-$,
$\phi p\bar p$, $2(\pi^+\pi^-)\eta$, $2(\pi^+\pi^-)\pi^0$,
$K^+K^-\pi^+\pi^-\pi^0$, $2(K^+K^-)\pi^0$, $p \bar p \pi^0$,
$p \bar p\pi^+\pi^-\pi^0$ and $3(\pi^+\pi^-)\pi^0$ produced in $e^+e^-$
annihilation at $\sqrt{s}=$ 3.773 and 3.650 GeV.
The measurements are made by analyzing the data sets of 17.3 pb$^{-1}$
collected at $\sqrt{s}=$ 3.773 GeV and 6.5 pb$^{-1}$ collected at
$\sqrt{s}=$ 3.650 GeV with the BESII detector at the BEPC collider.
For convenience, we call the data sets taken at $\sqrt{s}=$ 3.773 GeV and
$\sqrt{s}=$ 3.650 GeV to be the $\psi(3770)$ resonance data and the
continuum data in the Letter, respectively.

\section{BESII detector}
The BESII is a conventional cylindrical magnetic detector that is
described in detail in Refs. \cite {bes,bes2}. A 12-layer Vertex Chamber(VC)
surrounding a beryllium beam pipe provides input to event trigger,
as well as coordinate information. A forty-layer main drift chamber
(MDC) located just outside the VC yields precise measurements of charged
particle trajectories with a solid angle coverage of $85\%$ of 4$\pi$;
it also provides ionization energy loss ($dE/dx$) measurements for
particle identification. Momentum resolution of $1.7\%\sqrt{1+p^2}$
($p$ in GeV/$c$) and $dE/dx$ resolution of $8.5\%$ for Bhabha scattering
electrons are obtained for the data taken at $\sqrt{s}=$ 3.773 GeV. An
array of 48 scintillation counters surrounding the MDC measures time
of flight (TOF) of charged particles with a resolution of about 180
ps for electrons. Outside the TOF counters, a 12 radiation length,
lead-gas barrel shower counter (BSC), operating in limited streamer
mode, measures the energies of electrons and photons over $80\%$ of
the total solid angle with an energy resolution of $\sigma_E/E=0.22
/\sqrt{E}$ ($E$ in GeV) and spatial resolutions of $\sigma_{\phi}=7.9$
mrad and $\sigma_Z=2.3$ cm for electrons. A solenoidal magnet outside
the BSC provides a 0.4 T magnetic field in the central tracking region
of the detector. Three double-layer muon counters instrument the magnet
flux return and serve to identify muons with momentum greater than 500
MeV/c. They cover $68\%$ of the total solid angle.

\section{Event Selection}
\label{evtsel}
The exclusive light hadron final states mentioned above are reconstructed with the final
states $m(\pi^+\pi^-)n(K^+K^-)i(p\bar p) j(\gamma\gamma)\hspace{0.1cm}
(m=0,1,2,3;\hspace{0.1cm}n=0,1,2;\hspace{0.1cm}i=0,1\hspace{0.1cm}{\rm and}
\hspace{0.1cm}j=0,1)$, respectively. The $\phi$, $\pi^0$ and $\eta$
mesons are reconstructed through the decays $\phi\to K^+K^-$, $\pi^0\to
\gamma\gamma$ and $\eta\to\gamma\gamma$.

To select candidate events, it is first required that at least two
charged tracks be well reconstructed in the MDC with good helix fits. In
order to ensure the well-measured 3-momentum vectors and the reliability of
the charged-particle identification, the polar angle $\theta$ of each
charged track must satisfy $|\rm{cos\theta}|<0.85$. It is then required
that each charged track originate from the interaction region defined by
$\sqrt{V_x^2+V_y^2}<2.0$ cm and $|V_z|<20.0$ cm, where $V_x$, $V_y$ and
$V_z$ are the closest approach of the charged track in the $x$, $y$ and
$z$ directions.

Charged particles are identified by using the $dE/dx$ and TOF measurements,
with which the combined confidence levels ($CL_{\pi}$, $CL_K$ and $CL_p$)
for pion, kaon and proton hypotheses are calculated. A pion candidate is
required to have $CL_{\pi}> 0.001$. A kaon candidate is required to satisfy
$CL_K>CL_{\pi}$ in order to reduce misidentification. A proton candidate is
required to have the ratio $\frac{CL_p}{CL_{\pi}+CL_K+CL_p}$ greater than 0.6.

The photons are selected with the BSC measurements. It is required that
the energy deposited of the photon candidate in the BSC be greater than 50
MeV, and the electromagnetic shower start in the first 5 readout layers.
Moreover, the angle between the photon and the nearest charged track is
required to be greater than $22^\circ$ \cite{besdphy}, and the opening angle
between the direction of the shower development and the direction of the
photon emission is required to be less than $37^\circ$ \cite{besdphy}.

For each final state, there may be several different charged and/or neutral
track combinations satisfying the above selection criteria in an event.
In order to improve mass resolution, each combination is
subjected to an energy-momentum conservation kinematic fit. Only those
combinations with the kinematic fit probability greater than $1\%$ are
accepted. If more than one combination satisfies the selection criteria in
an event, only the combination with the largest fit probability is retained.

For the final state $2(\pi^+\pi^-)$, the background events of $(\gamma)J/\psi
\pi^+\pi^-$ with $J/\psi \to \mu^+\mu^-$ are vetoed by requiring that
the invariant mass of any $\pi^+\pi^-$ combination be less than 3.00
GeV/$c^2$.

For the final state $K^+K^-\pi^+\pi^-$ (or $K^+K^-\pi^+\pi^-\pi^0$),
the events from $D \bar D$ decays are excluded by rejecting the events
in which the $D^0$ and $\bar D^0$ mesons can be reconstructed in the decay
modes of $D^0\to K^-\pi^+$ and $\bar D^0\to K^+\pi^-$ (or $\bar D^0\to
K^+\pi^-\pi^0$, here charge conjugations are implied) \cite{besdcrs}.

\section{Data Analysis}
In the following data analysis, the modes containing $\phi$ and/or
$\pi^0/\eta$ mesons in the final states are studied by examining the
spectra of the $K^+K^-$ and/or $\gamma\gamma$ invariant masses ($M_{K^+K^-}$
and/or $M_{\gamma \gamma}$), which are calculated with the fitted momentum vectors
from the kinematic fit. In the $K^+K^-$ (or $\gamma\gamma$) invariant mass
spectrum for those final states with a few accepted events, the region within
the 20 MeV/$c^2$ (or $3\sigma^{\pi^0/\eta}_{M_{ \gamma\gamma}}$) window
around the $\phi$ (or $\pi^0/\eta$) nominal mass is taken as the $\phi$
(or $\pi^0/\eta$) signal region. Here, $\sigma^{\pi^0/\eta}_{M_{\gamma
\gamma}}$ is the standard deviation of the $\gamma\gamma$ invariant mass
spectrum determined by Monte Carlo simulation.

To study the modes without intermediate sub-resonance decays in the final
states, we define a kinematic quantity of
the scaled energy as $E_{\rm msr}/E_{\rm cm}$, where $E_{\rm msr}$ is
the total measured energy of all the observed particles, and $E_{\rm cm}$ is
the nominal center-of-mass energy.

\subsection{Candidates for $e^+e^-\to\phi\pi^0$ and $e^+e^-\to\phi\eta$}
\begin{figure}[htbp]
\begin{center}
\includegraphics*[width=8.0cm]
{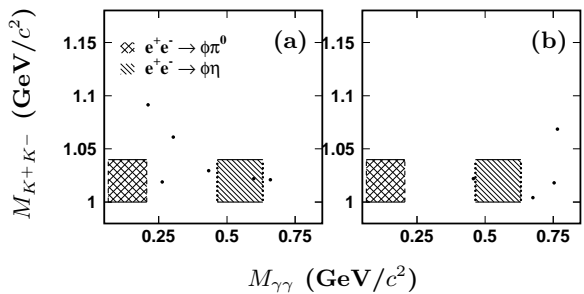}
\put(-145,0.0){\bf $M_{\gamma \gamma}$ (GeV/$c^2$)}
\put(-235,25.0){\rotatebox{90}{\bf $M_{K^+K^-}$ (GeV/$c^2$)}}
\put(-135.5,90){\bf (a)}
\put(-38,90){\bf (b)}
\caption{The scatter plots of $M_{K^+K^-}$ versus
$M_{\gamma \gamma}$ for the $K^+K^- \gamma \gamma$ combinations selected
from (a) the $\psi(3770)$ resonance data and (b) the continuum data, where the
hatched areas denote the $\phi\pi^0$ and $\phi\eta$ signal regions,
while the solid, dashed and dotted boundary lines represent the $\phi$,
$\pi^0$ and $\eta$ signal regions, respectively.}
\label{dot_kk_gg}
\end{center}
\end{figure}

The resulting scatter plots of $M_{K^+K^-}$ versus $M_{\gamma \gamma}$ for
the selected $K^+K^-\gamma\gamma$ events are shown in Fig. \ref{dot_kk_gg},
where the hatched areas show the $\phi\pi^0$ and $\phi
\eta$ signal regions, while the solid, dashed and dotted boundary lines
represent the $\phi$, $\pi^0$ and $\eta$ signal regions, respectively.
In the $\phi\pi^0$ ($\phi\eta$) signal region in the figure, there is
0(1) and 0(0) signal event for the $\phi\pi^0$ ($\phi\eta$) final state
observed from the $\psi(3770)$ resonance data and the continuum data,
respectively.

\subsection{Candidates for $e^+e^- \to m(\pi^+\pi^-)n(K^+K^-)i(p\bar p)$}
\begin{figure}[htbp]
\includegraphics*[width=8.0cm]
{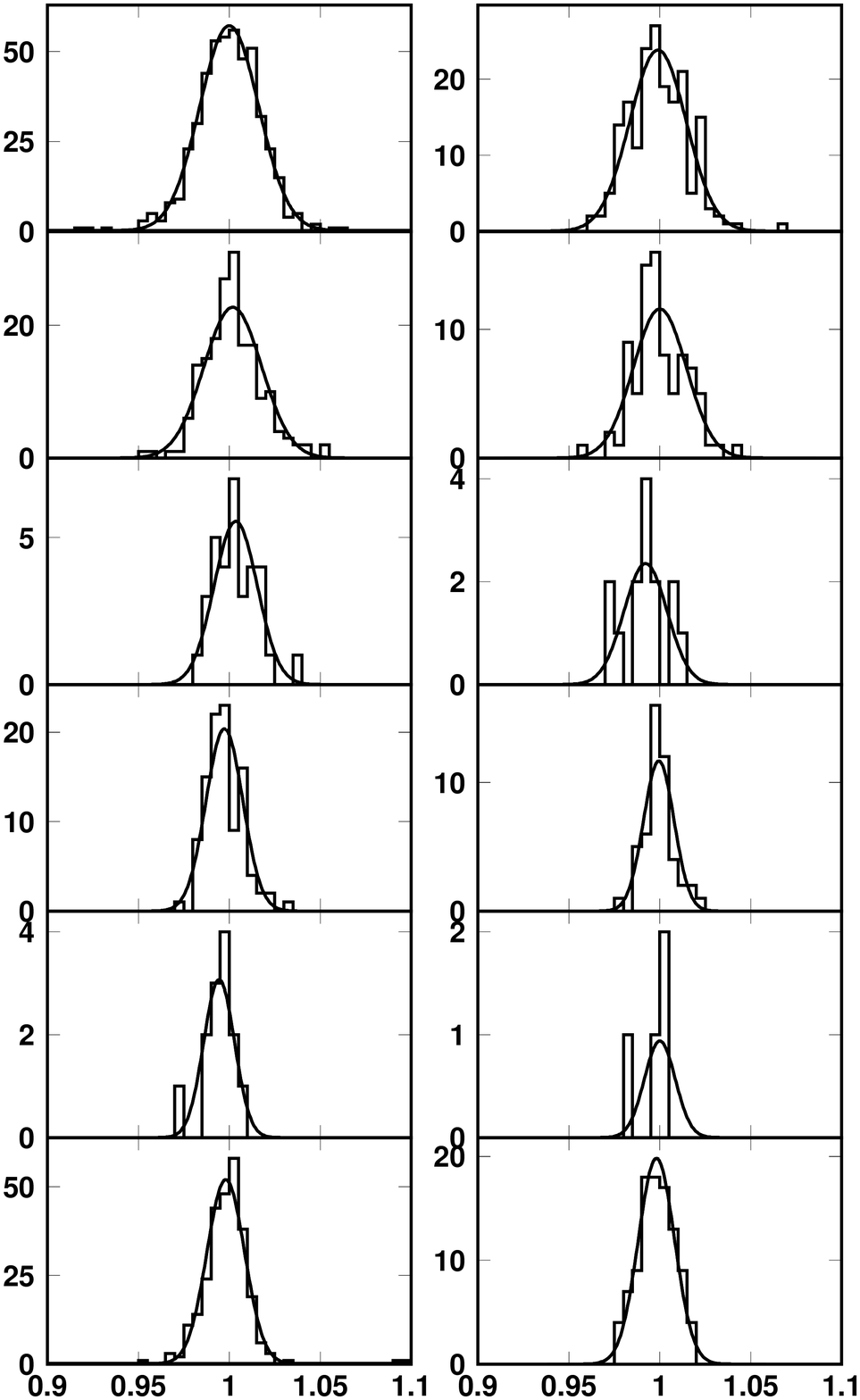}
\put(-125,0){\bf $E_{\rm msr}/E_{\rm cm}$}
\put(-235,140){\rotatebox{90}{\bf Events/(0.005)}}
\put(-135.5,317){\bf (a)}
\put(-135.5,266){\bf (b)}
\put(-135.5,215){\bf (c)}
\put(-135.5,164){\bf (d)}
\put(-135.5,113){\bf (e)}
\put(-135.5,62){\bf (f)}
\put(-38,317){\bf (a')}
\put(-38,266){\bf (b')}
\put(-38,215){\bf (c')}
\put(-38,164){\bf (d')}
\put(-38,113){\bf (e')}
\put(-38,62){\bf (f')}
\caption{
The $E_{\rm msr}/E_{\rm cm}$ distributions of the candidates for
(a) $e^+e^- \to 2(\pi^+\pi^-)$,
(b) $e^+e^- \to K^+K^-\pi^+\pi^-$,
(c) $e^+e^- \to 2(K^+K^-)$,
(d) $e^+e^- \to p \bar p \pi^+\pi^-$,
(e) $e^+e^- \to p \bar p K^+K^-$ and
(f) $e^+e^- \to 3(\pi^+\pi^-)$
selected from the $\psi(3770)$ resonance data (left) and the continuum data
(right).}
\label{ratiodata}
\end{figure}

Figure \ref{ratiodata} shows the resulting $E_{\rm msr}/E_{\rm cm}$
distributions, with each peak centered around unity correctly,
from the selected $m(\pi^+\pi^-)n(K^+K^-)i(p\bar p)$ events
for studying $e^+e^-\to 2(\pi^+\pi^-)$, $e^+e^-\to K^+K^-\pi^+\pi^-$,
$e^+e^-\to 2(K^+K^-)$, $e^+e^-\to p\bar p\pi^+\pi^-$, $e^+e^-\to p\bar p
K^+K^-$ and $e^+e^-\to 3(\pi^+\pi^-)$.
Fitting to the spectrum with a Gaussian function for the signal and a
flat background yields the number $N^{\rm obs}$ of the signal events
for each mode observed from the $\psi(3770)$ resonance data and the
continuum data.

\subsection{Candidates for $e^+e^- \to \phi K^+K^-$, $e^+e^- \to\phi\pi^+\pi^-$ and
$e^+e^- \to\phi p\bar p$}
\begin{figure}[htbp]
\begin{center}
\includegraphics*[width=8.0cm]
{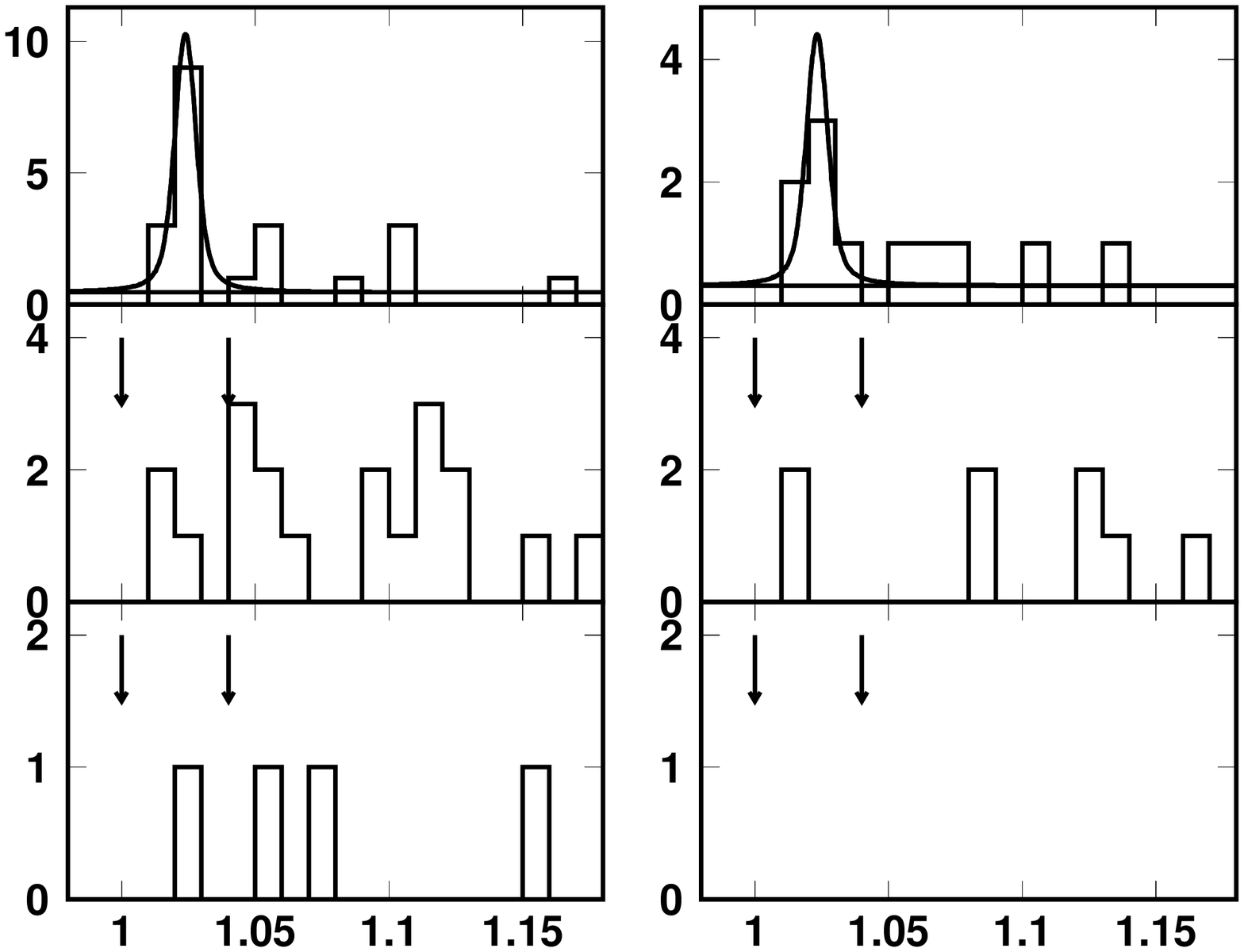}
\put(-145,0){\bf $M_{K^+K^-}$ (GeV/$c^2$)}
\put(-235,45){\rotatebox{90}{\bf Events/(0.01 GeV/$c^2$)}}
\put(-135.5,150){\bf (a)}
\put(-135.5,104){\bf (b)}
\put(-135.5,58){\bf (c)}
\put(-38,150){\bf (a')}
\put(-38,104){\bf (b')}
\put(-38,58){\bf (c')}
\caption{
The distributions of the invariant masses of the $K^+K^-$ combinations from
the candidates for
(a) $e^+e^- \to \phi K^+K^-$,
(b) $e^+e^- \to \phi \pi^+\pi^-$ and
(c) $e^+e^- \to \phi p \bar p$
selected from the $\psi(3770)$ resonance data (left) and the continuum data
(right), where the pairs of arrows represent the $\phi$ signal regions.}
\label{xmkk_phix}
\end{center}
\end{figure}

To investigate the processes $e^+e^-\to\phi K^+K^-$, $e^+e^-\to \phi\pi^+\pi^-$ and
$e^+e^-\to \phi p\bar p$, we further analyze the $K^+K^-$ invariant mass spectra of
the selected $2(K^+K^-)$, $K^+K^-\pi^+\pi^-$ and $p\bar p K^+K^-$ events,
as shown in Fig. \ref{xmkk_phix}. The $\phi$ signals can be seen in Figs.
\ref{xmkk_phix}(a) and \ref{xmkk_phix}(a'). Fitting to these two $K^+K^-$
invariant mass spectra with a Gaussian function for the $\phi$ signal and a
flat background yields $12.1\pm5.8$ and $4.9\pm2.6$ $\phi K^+K^-$ events
observed from the $\psi(3770)$ resonance data and the continuum data. In the
fits, the width of $\phi$ is fixed to 4.26 MeV/$c^2$ \cite{pdg}, and the
mass resolution is fixed to 3 MeV/$c^2$ determined by Monte Carlo
simulation. However, no obvious $\phi$ signals are observed in the other
figures. Selecting the events with $M_{K^+K^-}$ in the $\phi$ signal
regions, 
we obtain 3(1) and 2(0) signal events for the $\phi\pi^+\pi^-$ ($\phi p\bar p$)
final state from the $\psi(3770)$ resonance data and the continuum data, respectively.

\subsection{Candidates for $e^+e^- \to m(\pi^+\pi^-)
n(K^+K^-)i(p\bar p)\eta/\pi^0$}
\begin{figure}[htbp]
\begin{center}
\includegraphics*[width=8.0cm]
{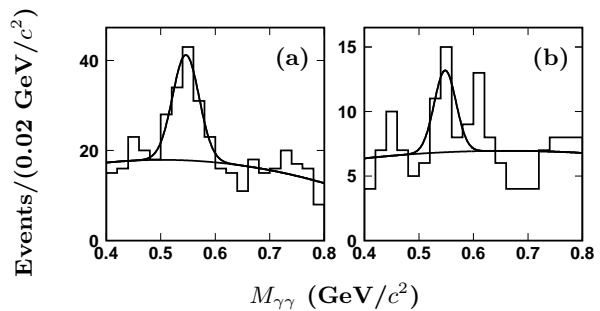}
\put(-145,0){\bf $M_{\gamma \gamma}$ (GeV/$c^2$)}
\put(-235,10){\rotatebox{90}{\bf Events/(0.02 GeV/$c^2$)}}
\put(-135.5,90){\bf (a)}
\put(-38,90){\bf (b)}
\caption{
The distributions of the invariant masses of the $\gamma\gamma$ combinations
from the candidates for $e^+e^- \to 2(\pi^+\pi^-) \eta$ selected from (a)
the $\psi(3770)$ resonance data and (b) the continuum data.}
\label{eta}
\end{center}
\end{figure}

\begin{figure}[htbp]
\includegraphics*[width=8.0cm]
{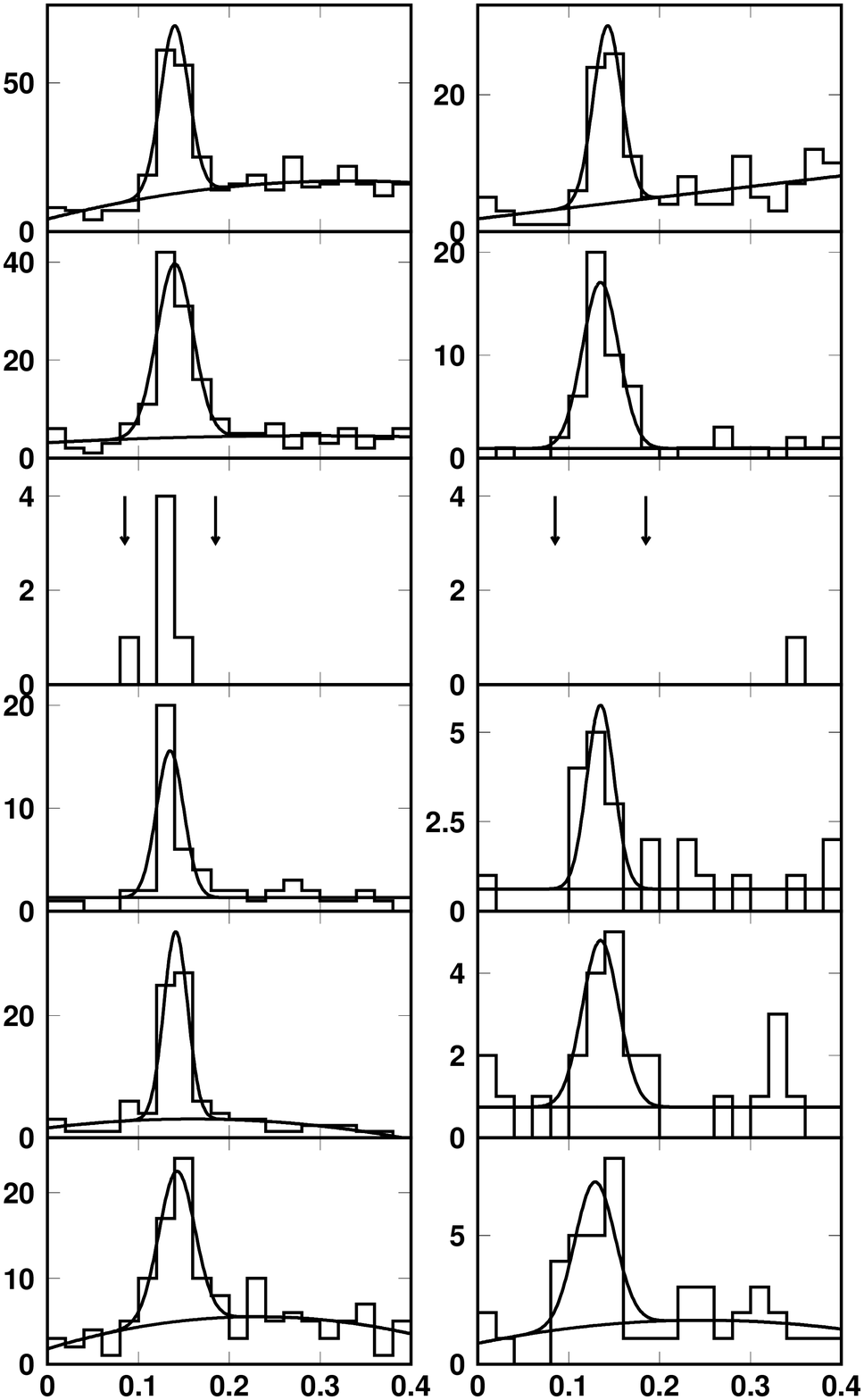}
\put(-145,0){\bf $M_{\gamma \gamma}$ (GeV/$c^2$)}
\put(-235,125){\rotatebox{90}{\bf Events/(0.02 GeV/$c^2$)}}
\put(-135.5,317){\bf (a)}
\put(-135.5,266){\bf (b)}
\put(-135.5,215){\bf (c)}
\put(-135.5,164){\bf (d)}
\put(-135.5,113){\bf (e)}
\put(-135.5,62){\bf (f)}
\put(-38,317){\bf (a')}
\put(-38,266){\bf (b')}
\put(-38,215){\bf (c')}
\put(-38,164){\bf (d')}
\put(-38,113){\bf (e')}
\put(-38,62){\bf (f')}
\caption{
The distributions of the invariant masses of the $\gamma\gamma$ combinations
from the candidates for
(a) $e^+e^- \to 2(\pi^+\pi^-)\pi^0$,
(b) $e^+e^- \to K^+K^-\pi^+\pi^-\pi^0$,
(c) $e^+e^- \to 2(K^+K^-)\pi^0$,
(d) $e^+e^- \to p \bar p \pi^0$,
(e) $e^+e^- \to p \bar p \pi^+\pi^-\pi^0$ and
(f) $e^+e^- \to 3(\pi^+\pi^-)\pi^0$
selected from the $\psi(3770)$ resonance data (left) and the continuum data (right),
where the pairs of arrows represent the $\pi^0$ signal regions.}
\label{pi0}
\end{figure}

Figures \ref{eta} and \ref{pi0} show the distributions of the $\gamma\gamma$
combinations from the selected $m(\pi^+\pi^-)n(K^+K^-)i(p\bar p)\gamma\gamma$
events for studying $e^+e^- \to 2(\pi^+\pi^-)\eta$, $e^+e^-\to 2(\pi^+
\pi^-)\pi^0$, $e^+e^-\to K^+K^-\pi^+\pi^-\pi^0$, $e^+e^-\to 2(K^+K^-)
\pi^0$, $e^+e^-\to p\bar p \pi^0$, $e^+e^-\to p \bar p \pi^+\pi^-\pi^0$
and $e^+e^-\to 3(\pi^+\pi^-)\pi^0$. In most of the figures, the
signals for $\eta/\pi^0$ production are clearly observed. Fitting to
these $\gamma\gamma$ invariant mass spectra with a Gaussian function
for the $\eta/\pi^0$ signal and a polynomial for the
background yields the number $N^{\rm obs}$ of the signal events for each mode
observed from the $\psi(3770)$ resonance data and the continuum data. In the fits to the
$\gamma\gamma$ invariant mass spectra in Figs. \ref{eta}(b), \ref{pi0}(d),
\ref{pi0}(d') and \ref{pi0}(e'), the mass and the mass resolution of $\eta/\pi^0$
are fixed to its nominal mass \cite{pdg} and 
$\sigma^{\eta/\pi^0}_{M_{\gamma\gamma}}$ determined by Monte Carlo
simulation, respectively.
However, there are only a few events in Figs. \ref{pi0}(c) and \ref{pi0}(c').
Selecting the events with $M_{\gamma\gamma}$ in the $\pi^0$ signal region,
we obtain 6 and 0 signal events for the $2(K^+K^-)\pi^0$ final state from the
$\psi(3770)$ resonance data and the continuum data, respectively.

\vspace{-1.0cm}
\section{Background Subtraction}
For some light hadron final states observed from the two data sets, there are the
contributions from $J/\psi$ and $\psi(3686)$ decays, the
contributions from the other final states due to misidentifying a $\pi$
as $K$ or reverse, and the contributions from $D\bar D$ decays. The
number of these contributions should be subtracted from the number $N^{\rm obs}$
of the candidate events for $e^+e^- \to f$ ($f$ represents light hadron final
state). These are estimated by Monte Carlo simulation.

\subsection{Contributions from $J/\psi$ decays}
The number of the contributions from the decay $J/\psi \to f$, where
$J/\psi$ is produced due to ISR returns, is
estimated by
\begin{equation}
N_{J/\psi \to f}
=\sigma_{J/\psi}
\times \mathcal{L} \times \epsilon_{J/\psi \to f}
\times {\mathcal B}(J/\psi \to f),
\label{eqjpsitail}
\end{equation}
where $\sigma_{J/\psi}$ is the cross section for $J/\psi$ production determined
with the $J/\psi$ resonance parameters \cite{pdg}, $\mathcal L$ is the integrated
luminosity of the data set, $\epsilon_{J/\psi \to f}$ is the detection
efficiency obtained by analyzing the Monte Carlo events of $J/\psi \to f$,
${\mathcal B}(J/\psi \to f)$ is the branching fraction for the decay
$J/\psi \to f$ \cite{pdg}.

\subsection{Contributions from $\psi(3686)$ decays}
The estimate of the contributions from the decay $\psi(3686)\to f$ is
similarly done with Eq. (\ref{eqjpsitail}) by substituting $J/\psi$ with
$\psi(3686)$. Here, the cross section $\sigma_{\psi(3686)}$ for $\psi(3686)$
production is determined with the $\psi(3686)$ resonance parameters measured
by BES Collaboration \cite{brdd2}.

The number of the contributions from the decay $\psi(3686) \to J/\psi\pi^+
\pi^-\to f$ is estimated by
\begin{eqnarray}
N_{\psi(3686) \to J/\psi \pi^+\pi^- \to f}
&=&
\sigma_{\psi(3686)} \times \mathcal{L}
\nonumber \\ & \times &
\epsilon_{\psi(3686) \to J/\psi\pi^+\pi^-,J/\psi \to f'}
\nonumber \\ & \times &
{\mathcal B}(\psi(3686) \to J/\psi \pi^+\pi^-)
\nonumber \\ & \times &
{\mathcal B}(J/\psi \to f'),
\label{eqpsitail2}
\end{eqnarray}
where
$\epsilon_{\psi(3686) \to J/\psi\pi^+\pi^-, J/\psi\to f'}$ is the
detection efficiency obtained by analyzing the Monte Carlo events
of $\psi(3686) \to J/\psi\pi^+\pi^-$ with $J/\psi \to f'$,
${\mathcal B}(\psi(3686) \to J/\psi \pi^+\pi^-)$ and
${\mathcal B}(J/\psi \to f')$ are the branching fractions for the decays
$\psi(3686) \to J/\psi\pi^+\pi^-$ and $J/\psi \to f'$ \cite{pdg},
respectively.

Monte Carlo study shows that the contributions from the decays
$\psi(3686) \to J/\psi\pi^0\pi^0$, $\psi(3686) \to J/\psi\pi^0$,
$\psi(3686) \to J/\psi\eta$, $\psi(3686) \to \gamma \chi_{cJ}
\hspace{0.1cm}(J=0,1,2)$ can be negligible.

\subsection{Cross background from the other final states}
The events from the other final state $e^+e^-\to i$ may be misidentified as the final
state $e^+e^-\to f$ in question. For example, the final states $K^+K^-\pi^+\pi^-$ and
$2(\pi^+\pi^-)\pi^0$ may be faked as $2(\pi^+\pi^-)$ due to misidentifying
$K^+K^-$ pair as $\pi^+\pi^-$ pair and due to missing $\pi^0$.
The number of these background events from $e^+e^-\to i$ to $e^+e^-\to f$ is
estimated by
\begin{equation}
N_{\rm CB}
=\sigma_{e^+e^- \to i} \times \mathcal{L} \times \epsilon_{i \to f},
\label{crsback}
\end{equation}
where $\sigma_{e^+e^- \to i}$ is the observed cross section for the final
state $e^+e^- \to i$, $\epsilon_{i \to f}$ is the rate of misidentifying
$e^+e^- \to i$ as $e^+e^- \to f$.

\subsection{Contributions from $D \bar D$ decays}
In the $\psi(3770)$ resonance region, the events from $D \bar D$ decays may
also satisfy the selection criteria for the light hadron final states. The
number of these events is estimated by analyzing a Monte Carlo sample which
is about 78 times larger than the $\psi(3770)$ resonance data. The Monte Carlo events are
generated as $e^+e^- \to D\bar D$, where the $D$ and $\bar D$ mesons are
set to decay into all possible final states with the branching fractions
quoted from PDG \cite{pdg}. The number of events satisfying the selection
criteria for $e^+e^- \to f$ in question is then normalized to the data,
resulting in the number $N_{D\bar D}$ of the contributions
from $D\bar D$ decays to $e^+e^- \to f$.

Monte Carlo study shows that the contributions from the decays $\psi(3770)
\to J/\psi\pi^+\pi^-$, $\psi(3770) \to J/\psi\pi^0\pi^0$, $\psi(3770) \to
J/\psi\pi^0$, $\psi(3770) \to J/\psi\eta$, $\psi(3770) \to\gamma \chi_{cJ}
\hspace{0.1cm}(J=0,1,2)$ can be negligible.

\subsection{Number of signal events}
Adding $N_{J/\psi}$, $N_{\psi(3686)}$, $N_{\rm CB}$ and $N_{D\bar D}$
yields the total number $N^{\rm b}$ of the background events for each final
state.
After subtracting $N^{\rm b}$ from $N^{\rm obs}$, we obtain the numbers
$N^{\rm net}$ of the signal events for some final states, under the assumption
that there is no interference between the light hadron final states of the
resonance decays and those from the continuum production. For the other
final states with a few events observed from the data, we set the upper
limits $N^{\rm up}$ on these numbers of the signal events by using the Feldman-Cousins
method \cite{felman} in the absence of background at 90\% confidence level
(C.L.). The second, third and fourth columns of Tabs \ref{crs3773} and
\ref{crs3650} summarize the results on the numbers $N^{\rm obs}$, $N^{\rm
b}$ and $N^{\rm net}$ (or $N^{\rm up}$).

\section{Results}
\subsection{Monte Carlo efficiency}
The detection efficiency $\epsilon$ for each final state is estimated by
Monte Carlo simulation for the BESII detector \cite{bessim}. The Monte Carlo
events are generated by using a phase space generator including initial
state radiation and photon vacuum polarization
corrections \cite{isr} with $1/s$ cross section energy dependence.
Final state radiation \cite{fsr} decrease the detection efficiency not
more than 0.5\%.
Detailed analysis gives the detection
efficiency for each final state at $\sqrt{s}=$ 3.773 and 3.650 GeV.
They are summarized in the fifth columns of Tabs. \ref{crs3773} and
\ref{crs3650}. For the modes containing $\phi$, $\pi^0$ and $\eta$ mesons
in the final states, the detection efficiencies listed in the tables do not include the
branching fractions for the decays $\phi\to K^+K^-$, $\pi^0\to\gamma\gamma$
and $\eta \to\gamma\gamma$.

\subsection{Systematic error}
\label{sys}
In the measurements of the observed cross sections for $e^+e^-\to$ light
hadrons, the systematic error
arises mainly from the uncertainties in luminosity ($\sim$2.1\%
\cite{brdd1,brdd2}), photon selection ($\sim$2.0\% per photon), tracking
efficiency ($\sim$2.0\% per track), particle identification ($\sim$0.5\%
per pion or kaon, $\sim$2.0\% per proton), kinematic fit ($\sim$1.5\%),
Monte Carlo statistics ($\sim$(1.1$\sim$3.2)\%), branching fractions quoted
from PDG \cite{pdg} ($\sim$1.22\% for $\mathcal B(\phi \to K^+K^-)$,
$\sim$0.03\% for $\mathcal B(\pi^0\to\gamma\gamma)$ and $\sim$0.66\% for
$\mathcal B(\eta\to\gamma\gamma)$), background subtraction
($\sim$(0.0$\sim$4.8)\%), fit to mass spectrum ($\sim$3.8\%),
and Monte Carlo modeling ($\sim$6.0\%).
Adding these uncertainties in quadrature
yields the total systematic error $\Delta_{\rm sys}$ for each final state at
$\sqrt{s}=$ 3.773 and 3.650 GeV, respectively, which are shown in the sixth
columns of Tabs. \ref{crs3773} and \ref{crs3650}.

\subsection{Observed cross sections for some final states}
The observed cross section for $e^+e^- \to f$ is determined by
\begin{equation}
\sigma = \frac{N^{\rm net}} {\mathcal{L}
\times \epsilon
[\times {\mathcal B}(\phi \to K^+K^-)]
[\times {\mathcal B}(\pi^0/\eta \to \gamma\gamma)]},
\label{crs}
\end{equation}
where $N^{\rm net}$ is the number of the signal events, $\epsilon$ is the
detection efficiency for the final sate, and
${\mathcal B}(\phi \to K^+K^-)$ and ${\mathcal B}(\pi^0/\eta \to\gamma\gamma)$
are the branching fractions for the decays $\phi\to K^+K^-$ and $\pi^0/\eta \to \gamma
\gamma$ \cite{pdg}, respectively. Inserting these numbers in Eq. (\ref{crs}), we obtain the
observed cross sections for some final states at $\sqrt{s}=$ 3.773 and 3.650
GeV, respectively. They are shown in the last
columns of Tabs. \ref{crs3773} and \ref{crs3650}, where the
first error is statistical and the second systematic. The measured observed
cross sections for $e^+e^- \to 3(\pi^+\pi^-)$ and $e^+e^- \to K^+K^-\pi^+\pi^-\pi^0$ at
$\sqrt{s}=$ 3.650 GeV are consistent with the results in Refs.
\cite{6pi,2k3pi} within error.

\subsection{Upper limits on the observed cross sections for the other final
states}
The upper limit on the observed cross section for $e^+e^- \to f$ with a few
events observed from the data is set by
\begin{eqnarray}
&&\sigma^{\rm up}=
\nonumber \\ &&
\hspace{-0.5cm}
\frac{N^{\rm up}} {\mathcal{L} \times
\epsilon
\times (1-\Delta_{\rm sys})
[\times {\mathcal B}(\phi \to K^+K^-)]
[\times {\mathcal B}(\pi^0/\eta \to \gamma\gamma)]},
\nonumber \\ &&
\label{crsup}
\end{eqnarray}
where $N^{\rm up}$ is the upper limit on the number of the signal events, and
$\Delta_{\rm sys}$ is the systematic error in the measurement of the
observed cross section for the final state. Inserting
the corresponding numbers in Eq. (\ref{crsup}), we obtain the upper limits on
the observed cross sections for these final states at $\sqrt{s}=$ 3.773 and
3.650 GeV, respectively, which are also shown in the last columns of Tabs.
\ref{crs3773} and \ref{crs3650}. The upper limits on the observed cross
sections for $e^+e^- \to\phi\pi^0$ and $e^+e^- \to\phi\eta$ are consistent with
the measurements \cite{adams} from the CLEO Collaboration.

\begin{table*}[hbtp]
\begin{center}
\caption{
The observed cross sections for $e^+e^-\to$ light hadrons at $\sqrt{s}=$ 3.773 GeV, where
$N^{\rm obs}$ is the number of events observed from the $\psi(3770)$
resonance data, $N^{\rm b}$ is the total number of background events,
$N^{\rm net}$ is the number of the signal events,
$N^{\rm up}$ is the upper limit on the number of the signal events,
$\epsilon$ is the detection efficiency for the final state,
$\Delta_{\rm sys}$ is the relative systematic error in the measurement of
the observed cross section,
$\sigma$ is the observed cross section and
$\sigma^{\rm up}$ is the upper limit on the observed cross section set at 90\% C.L..}
\begin{tabular}{|l|c|c|c|c|c|c|} \hline
$e^+e^-\to$ &$N^{\rm obs}$&$N^{\rm b}$
&$N^{\rm net}$ (or $N^{\rm up}$)
&$\epsilon$[\%]&$\Delta_{\rm sys}$[\%]
&$\sigma$ (or $\sigma^{\rm up}$) [pb] \\ \hline
$\phi\pi^0$             &0             &0           &$<2.44$       &$ 9.08\pm0.14$& 8.9&$<3.5$               \\
$\phi\eta$              &1             &0           &$<4.36$       &$11.31\pm0.16$& 8.9&$<12.6$              \\
$2(\pi^+\pi^-)$         &$462.5\pm21.9$&$7.8 \pm1.0$&$454.7\pm21.9$&$15.13\pm0.18$&10.6&$173.7\pm8.4 \pm18.4$\\
$K^+K^-\pi^+\pi^-$      &$181.0\pm13.5$&$4.9 \pm0.6$&$176.1\pm13.5$&$ 7.73\pm0.13$&10.7&$131.7\pm10.1\pm14.1$\\
$\phi\pi^+\pi^-$        &3             &0           &$<7.42$       &$ 8.81\pm0.13$&10.7&$<11.1$              \\
$2(K^+K^-)$             &$33.0\pm5.8$  &$1.1 \pm0.2$&$31.9 \pm5.8$ &$ 9.26\pm0.14$&10.6&$19.9 \pm3.6 \pm2.1$ \\
$\phi K^+K^-$           &$12.1\pm3.9$  &$0.1 \pm0.0$&$12.0 \pm3.9$ &$ 8.94\pm0.14$&11.4&$15.8 \pm5.1 \pm1.8$ \\
$p \bar p \pi^+\pi^-$   &$103.0\pm10.2$&$4.0 \pm0.4$&$99.0 \pm10.2$&$17.26\pm0.19$&11.5&$33.2 \pm3.4 \pm3.8$ \\
$p \bar p K^+K^-$       &$13.0\pm3.7$  &$0.2 \pm0.0$&$12.8 \pm3.7$ &$10.48\pm0.14$&11.6&$7.1  \pm2.0 \pm0.8$ \\
$\phi p \bar p$         &1             &0           &$<4.36$       &$ 9.96\pm0.14$&11.6&$<5.8$               \\
$3(\pi^+\pi^-)$         &$268.7\pm16.5$&$3.8 \pm0.6$&$264.9\pm16.5$&$ 6.47\pm0.11$&14.1&$236.7\pm14.7\pm33.4$\\
$2(\pi^+\pi^-)\eta$     &$71.4\pm18.6$ &$0.1 \pm0.1$&$71.3 \pm18.6$&$ 6.81\pm0.12$&12.0&$153.7\pm40.1\pm18.4$\\
$2(\pi^+\pi^-)\pi^0$    &$110.2\pm14.4$&$26.5\pm2.7$&$83.7 \pm14.4$&$ 6.05\pm0.11$&12.4&$80.9 \pm13.9\pm10.0$\\
$K^+K^-\pi^+\pi^-\pi^0$ &$88.3\pm12.6$ &$5.3 \pm0.7$&$83.0 \pm12.6$&$ 2.83\pm0.08$&12.2&$171.6\pm26.0\pm20.9$\\
$2(K^+K^-)\pi^0$        &$6.0\pm2.5$   &$0.1 \pm0.1$&$ 5.9 \pm2.5$ &$ 1.91\pm0.06$&11.8&$18.1 \pm7.7 \pm2.1$ \\
$p\bar p\pi^0$          &$26.8\pm5.7$  &$0.7 \pm0.1$&$26.1 \pm5.7$ &$15.12\pm0.18$&10.3&$10.1 \pm2.2 \pm1.0$ \\
$p\bar p\pi^+\pi^-\pi^0$&$50.4\pm8.3$  &$2.3 \pm0.4$&$48.1 \pm8.3$ &$ 5.30\pm0.11$&12.9&$53.1 \pm9.2 \pm6.8$ \\
$3(\pi^+\pi^-)\pi^0$    &$43.6\pm10.1$ &$12.5\pm1.5$&$31.1 \pm10.1$&$ 1.72\pm0.04$&16.0&$105.8\pm34.4\pm16.9$\\
\hline
\end{tabular}
\label{crs3773}
\end{center}
\end{table*}

\begin{table*}[hbtp]
\begin{center}
\caption{
The observed cross sections for $e^+e^-\to$ light hadrons at $\sqrt{s}=$ 3.650 GeV, where
$N^{\rm obs}$ is the number of events observed from the continuum data,
and the definitions of the other symbols are the same as those in Tab. \ref{crs3773}.}
\begin{tabular}{|l|c|c|c|c|c|c|} \hline
$e^+e^-\to$ &$N^{\rm obs}$&$N^{\rm b}$
&$N^{\rm net}$ (or $N^{\rm up}$)
&$\epsilon$[\%]&$\Delta_{\rm sys}$[\%]
&$\sigma$ (or $\sigma^{\rm up}$) [pb] \\ \hline
$\phi\pi^0$             &0             &0          &$<2.44$       &$9.48 \pm0.14$& 8.9&$<8.9$               \\
$\phi\eta$              &0             &0          &$<2.44$       &$11.82\pm0.16$& 8.9&$<18.0$              \\
$2(\pi^+\pi^-)$         &$185.2\pm13.7$&$1.9\pm0.3$&$183.3\pm13.7$&$15.87\pm0.18$&10.6&$177.7\pm13.3\pm18.8$\\
$K^+K^-\pi^+\pi^-$      &$85.0 \pm 9.3$&$0.9\pm0.1$&$84.1 \pm 9.3$&$8.00 \pm0.13$&10.6&$161.7\pm17.9\pm17.1$\\
$\phi\pi^+\pi^-$        &2             &0          &$<5.91$       &$9.08 \pm0.13$&10.7&$<22.9$              \\
$2(K^+K^-)$             &$14.0 \pm 3.8$&0          &$14.0 \pm 3.8$&$8.94 \pm0.13$&10.6&$24.1 \pm6.5 \pm 2.6$\\
$\phi K^+K^-$           &$4.9  \pm 2.6$&0          &$ 4.9 \pm 2.6$&$8.81 \pm0.14$&11.4&$17.4 \pm9.2 \pm 2.0$\\
$p \bar p \pi^+\pi^-$   &$49.0 \pm 7.0$&$0.8\pm0.1$&$48.2 \pm 7.0$&$17.60\pm0.19$&11.5&$42.1 \pm6.1 \pm 4.8$\\
$p \bar p K^+K^-$       &$4.0  \pm 2.0$&0          &$ 4.0 \pm 2.0$&$10.03\pm0.14$&11.6&$6.1  \pm3.1 \pm 0.7$\\
$\phi p \bar p$         &0             &0          &$<2.44$       &$9.49 \pm0.14$&11.6&$<9.1$               \\
$3(\pi^+\pi^-)$         &$99.0 \pm10.0$&$0.2\pm0.0$&$98.8 \pm10.0$&$6.47 \pm0.11$&14.1&$234.9\pm23.8\pm33.1$\\
$2(\pi^+\pi^-)\eta$     &$15.9 \pm 7.4$&0          &$15.9 \pm 7.4$&$7.17 \pm0.13$&12.0&$86.6 \pm40.3\pm10.4$\\
$2(\pi^+\pi^-)\pi^0$    &$51.0 \pm 8.5$&$2.3\pm0.4$&$48.7 \pm 8.5$&$6.10 \pm0.11$&12.0&$124.3\pm21.7\pm14.9$\\
$K^+K^-\pi^+\pi^-\pi^0$ &$40.5 \pm 6.8$&$0.3\pm0.1$&$40.2 \pm 6.8$&$2.81 \pm0.08$&12.2&$222.8\pm37.7\pm27.2$\\
$2(K^+K^-)\pi^0$        &0             &0          &$<2.44$       &$1.87 \pm0.06$&11.7&$<23.0$              \\
$p\bar p\pi^0$          &$9.7  \pm 3.5$&$0.1\pm0.0$&$ 9.6 \pm 3.5$&$16.19\pm0.19$&10.3&$9.2  \pm3.4 \pm 1.0$\\
$p\bar p\pi^+\pi^-\pi^0$&$10.1 \pm 3.8$&$0.2\pm0.1$&$ 9.9 \pm 3.8$&$5.31 \pm0.11$&12.9&$29.0 \pm11.1\pm 3.7$\\
$3(\pi^+\pi^-)\pi^0$    &$15.5 \pm 5.5$&$0.7\pm0.1$&$14.8 \pm 5.5$&$1.82 \pm0.04$&15.2&$126.6\pm47.1\pm19.2$\\
\hline
\end{tabular}
\label{crs3650}
\end{center}
\end{table*}

\subsection{Upper limits on the observed cross section and the
branching fraction for $\psi(3770)\to f$}
If we don't consider the interference effects between the continuum and
resonance amplitudes, we may calculate the observed cross section
for $\psi(3770)\to f$ by
\begin{equation}
\sigma_{\psi(3770)\to f}=
\sigma^{\rm 3.773\hspace{0.05cm}GeV}_{e^+e^-\to f} -
f_{\rm co}\times \sigma^{\rm 3.650\hspace{0.05cm}GeV}_{e^+e^- \to f},
\label{obscrs}
\end{equation}
where $\sigma^{\rm 3.773\hspace{0.05cm}GeV}_{e^+e^- \to f}$ and
$\sigma^{\rm 3.650 \hspace{0.05cm}GeV}_{e^+e^- \to f}$ are the
measured observed cross sections for $e^+e^- \to f$ at
3.773 and 3.650 GeV, respectively, and $f_{\rm co}$ is the coefficient
due to the 1/s dependence of the cross section.
The results on $\sigma_{\psi(3770)\to f}$ are shown in the
second column of Tab. \ref{uppsipp}, where the first error is statistical,
the second is the independent systematic, and the third is the common
systematic error. Here, the independent systematic error is from the
uncertainties in the Monte Carlo statistics, in fitting to the mass spectrum
and in the background subtraction,
while the common systematic error is from the other uncertainties as
discussed in the subsection B.
For the final states of $\phi\pi^0$, $\phi\eta$, $\phi\pi^+\pi^-$,
$\phi p\bar p$ and $2(K^+K^-)\pi^0$, only a few signal events are observed
from both the $\psi(3770)$ resonance data and the continuum data,
therefore we neglect their contributions from the continuum production.

\begin{table*}[htbp]
\begin{center}
\caption{
The upper limits on the observed cross section $\sigma^{\rm up}_{\psi(3770)\to f}$
and the branching fraction ${\mathcal B}^{\rm up}_{\psi(3770)\to f}$ for
$\psi(3770)\to f$ are set at 90\% C.L.. The $\sigma_{\psi(3770)\to f}$ in
the second column is calculated with Eq. (\ref{obscrs}), where the first error
is the statistical, the second is the independent systematic, and the third is
the common systematic error. Here, the upper $^t$
denotes that we treat the upper limit on the observed cross section for
$e^+e^-\to f$ at 3.773 GeV as $\sigma^{\rm up}_{\psi(3770)\to f}$,
the upper $^n$ denotes that we neglect the
contribution from the continuum production, and the upper $^z$ denotes that we treat
the central value of $\sigma_{\psi(3770)\to f}$ as zero if it is less
than zero.}
\begin{tabular}{|l|c|c|c|} \hline
Decay Mode
   & $\sigma_{\psi(3770)\to f}$
   & $\sigma^{\rm up}_{\psi(3770)\to f}$
   & ${\mathcal B}^{\rm up}_{\psi(3770)\to f}$ \\
   & [pb] & [pb] & [$\times 10^{-3}$] \\ \hline
$\phi\pi^0$             &$< 3.5^{tn}$                   &$<  3.5$&$< 0.5$\\
$\phi\eta$              &$<12.6^{tn}$                   &$< 12.6$&$< 1.9$\\
$2(\pi^+\pi^-)$         &$  7.4\pm15.0\pm 2.8\pm0.8$    &$< 32.5$&$< 4.8$\\
$K^+K^-\pi^+\pi^-$      &$-19.6\pm19.6\pm 3.3\pm2.1^{z}$&$< 32.7$&$< 4.8$\\
$\phi\pi^+\pi^-$        &$<11.1^{tn}$                   &$< 11.1$&$< 1.6$\\
$2(K^+K^-)$             &$ -2.7\pm 7.1\pm 0.5\pm0.3^{z}$&$< 11.6$&$< 1.7$\\
$\phi K^+K^-$           &$ -0.5\pm10.0\pm 0.9\pm0.1^{z}$&$< 16.5$&$< 2.4$\\
$p \bar p \pi^+\pi^-$   &$ -6.2\pm 6.6\pm 0.6\pm0.7^{z}$&$< 11.0$&$< 1.6$\\
$p \bar p K^+K^-$       &$  1.4\pm 3.5\pm 0.1\pm0.2$    &$<  7.2$&$< 1.1$\\
$\phi p \bar p$         &$< 5.8^{tn}$                   &$<  5.8$&$< 0.9$\\
$3(\pi^+\pi^-)$         &$ 16.9\pm26.7\pm 5.5\pm2.4$    &$< 61.7$&$< 9.1$\\
$2(\pi^+\pi^-)\eta$     &$ 72.7\pm55.0\pm 7.3\pm8.2$    &$<164.7$&$<24.3$\\
$2(\pi^+\pi^-)\pi^0$    &$-35.4\pm24.6\pm 6.6\pm4.0^{z}$&$< 42.3$&$< 6.2$\\
$K^+K^-\pi^+\pi^-\pi^0$ &$-36.9\pm43.8\pm12.8\pm4.2^{z}$&$< 75.2$&$<11.1$\\
$2(K^+K^-)\pi^0$        &$ 18.1\pm 7.7\pm 0.7\pm2.0^{n}$&$< 31.2$&$< 4.6$\\
$p\bar p\pi^0$          &$  1.5\pm 3.9\pm 0.5\pm0.1$    &$<  7.9$&$< 1.2$\\
$p\bar p\pi^+\pi^-\pi^0$&$ 26.0\pm13.9\pm 2.6\pm3.2$    &$< 49.7$&$< 7.3$\\
$3(\pi^+\pi^-)\pi^0$    &$-12.7\pm55.9\pm 8.7\pm1.8^{z}$&$< 92.8$&$<13.7$\\ \hline
\end{tabular}
\label{uppsipp}
\end{center}
\end{table*}

The upper limits on the observed cross sections for $\psi(3770)$
decay to $\phi\pi^0$, $\phi\eta$, $\phi\pi^+\pi^-$ and
$\phi p\bar p$ are directly set based on the upper limits on their
observed cross sections at 3.773 GeV.
However, the upper limits on the observed cross sections for
$\psi(3770)$ decay to the other final states are set by shifting the cross
section by $1.64\sigma$, where the $\sigma$ is the total error of the
measured cross section.
If the central value of $\sigma_{\psi(3770)\to f}$ is less than zero,
we treat it as zero, then we set its upper limit.
The third column of Tab. \ref{uppsipp} summarizes the results on
$\sigma^{\rm up}_{\psi(3770) \to f}$.

BES Collaboration measured the observed cross section
for the $\psi(3770)$ production at $\sqrt {s}=$ 3.773 GeV
to be $6.94\pm0.48\pm0.28$ nb \cite{brdd2}
and $7.25\pm0.27\pm0.34$ nb \cite{rval} with two different data samples.
Averaging these two results, we obtain the weighted average cross section
$\sigma^{\rm obs}_{\psi(3770)}=7.15\pm0.27\pm0.27$ nb, where the first error is the combined
statistical and independent systematic error, and the second is the common systematic
error of the two measurements. The upper limit on the branching fraction
${\mathcal B}^{\rm up}_{\psi(3770)\to f}$ for
$\psi(3770)\to f$ is set by dividing $\sigma^{\rm up}_{\psi(3770) \to
f}$ by $\sigma^{\rm obs}_{\psi(3770)}$.
The fourth column of Tab. \ref{uppsipp} summarizes the results on
${\mathcal B}^{\rm up}_{\psi(3770)\to f}$.

\vspace{-0.5cm}
\section{Summary}
In summary, using the data sets collected at $\sqrt{s}=$ 3.773 and 3.650 GeV
with the BESII detector at the BEPC collider, we have measured the observed
cross sections for 18 exclusive light hadron final states produced in
$e^+e^-$ annihilation at the two energy points. These cross sections
exclude the contributions from $J/\psi$, $\psi(3686)$ and $D\bar D$
decays. The initial state radiation and photon vacuum polarization
corrections are taken into account in the determination of the detection
efficiency. In the measurements
we do not consider the interference effects between the continuum and
resonance amplitudes, since we do not know the details about the
two amplitudes. For this reason, although we do not observe significant
difference between the observed cross sections for most light hadron final
states at the two energy points, we can not draw a conclusion that the
$\psi(3770)$ does not decay into these final states. To extract the
branching fractions for $\psi(3770)\to$ light hadrons from the observed
cross sections, one needs to collect large data samples at more different
energy points, i.e. cross section scan, covering both $\psi(3770)$ and
$\psi(3686)$ \cite{crsscan}. This will
hopefully be done with the coming BESIII detector at the BEPCII collider in the
near future. However, the measurements of the observed cross sections for
these light hadron final states at the two energy points provide useful
information to understand the mechanisms of the continuum light hadron
production and still provide valuable information to understand the
discrepancy between the observed cross sections for $D\bar D$
and $\psi(3770)$ production.

\section{acknowledgments}
The BES collaboration thanks the staff of BEPC for their hard
efforts. This work is supported in part by the National Natural
Science Foundation of China under contracts Nos. 10491300,
10225524, 10225525, 10425523, the Chinese Academy of Sciences
under contract No. KJ 95T-03, the 100 Talents Program of CAS under
Contract Nos. U-11, U-24, U-25, the Knowledge Innovation Project
of CAS under Contract Nos. U-602, U-34 (IHEP), the National
Natural Science Foundation of China under Contract  No. 10225522
(Tsinghua University).

\end{document}